%% file: main.tex
\documentclass[orivec,runningheads]{llncs}
\usepackage{amsmath}
\DeclareMathOperator{\encode}{encode}
\usepackage[T1]{fontenc}
\usepackage{graphicx}
\usepackage{comment}
\usepackage{booktabs}
\usepackage{amsmath}
\usepackage{array}
\usepackage{xspace}
\usepackage{url}
\usepackage{tikz}
\usepackage{float}
\newcommand{\spacetune}[1]{#1}

\title{\mbox{Look Before You Leap:}  \mbox{Checking In on Type Tag Checking}}
\titlerunning{Look Before You Leap}

\author{Stephen M. Watt}

\institute{
Cheriton School of Computer Science\\
University of Waterloo\\ Waterloo, Ontario, Canada\\
\email{smwatt@uwaterloo.ca}
}

\begin{document}

\maketitle

\begin{abstract}
Tagging of generic dynamic values is important in symbolic-computation
and dynamic-language systems, but the trade-offs change as machine
architectures and workloads evolve.  In particular, old folklore about
boxed values, immediate values, and type tags must be recalibrated from
time to time.
We revisit the performance of badged object headers, low-bit 
tagging, and two NaN-boxing layouts on a range of platforms in use
today, including AArch64 and x86-64 architectures from different
manufacturers.  The experiments isolate two distinct effects: the cost
avoided by not heap-allocating common scalar values, and the cost
avoided by obtaining tag information from the value word rather than by
performing a heap read.  The results show that several local bit
operations are often cheaper than opening a heap object to obtain a tag
or small value.  Low-bit tagging remains the simplest and usually
fastest choice for mostly symbolic workloads, while NaN-boxing is close
in access cost and avoids the time and space of heap allocation for
ordinary floating-point values.

\end{abstract}

\section{Introduction}
Many programming languages require data values to carry dynamic type information. These include important mainstream platforms such as JavaScript and Python, as well as Lisp systems and computer algebra systems. In such systems, common small values such as fixed-size integers, truth values and characters may occur in very large numbers and may be inspected repeatedly inside generic code. Representing these directly in machine words, rather than as separately allocated heap objects, can avoid allocation, reduce memory traffic, and make shallow type tests cheaper.

This representation question now arises in a broader setting beyond traditional Lisp-like symbolic environments. Widely used dynamic languages carry numeric values within generic representations, and symbolic-numeric algorithms have brought algebraic and floating-point values into closer contact \cite{ICMS2026SymbolicNumeric,SNC-TCS}. A variety of schemes have historically been used to fit immediate scalar values and heap-object references into a single machine word, while still allowing the runtime to distinguish them dynamically \cite{Goldberg1991,Gudeman1993,IEEE7542019,MelanconSerranoFeeley2025}. Representation choices once judged mainly for symbolic systems must therefore be reconsidered for modern systems.

Even with as much static checking and specialization as a compiler can
safely add, values crossing generic interfaces may need to be
dynamically identifiable.  The practical question is therefore not
whether dynamic checks can be eliminated, but how cheaply a runtime can
represent and specialize the value classes that dominate execution.
For this reason, runtime type testing has long been a central
implementation concern in Lisp systems, dynamic language runtimes, and
computer algebra systems
\cite{Gabriel1985,Queinnec1996,Steele1977Data,Steele1977Fast,Steele1990}.

The question is old, but the answer is not eternal.  The relative costs
of integer operations, dependent memory loads, cache misses, branch
mispredictions, and compiler transformations have changed repeatedly
over the history of these software systems.  It is therefore risky to
rely on representation folklore formed from earlier machines.  
The purpose of this paper is 
to re-evaluate 
implementation trade-off under current conditions.

We compare three strategies.  In the first, \textit{``badged object
headers''}, all objects are heap-allocated and their type is recorded in a
header word.  
This is simple, uniform,
and extensible, but it requires reading memory.  In the second strategy,
\textit{``low-bit pointer tagging''}, low-order bits of aligned pointers
and shifted integers are used to encode type information.  This avoids
a header access for common cases but imposes masking, shifting, and
reconstruction operations.  In the third strategy, \textit{``NaN
boxing''}, values are encoded using the large space of non-numeric IEEE
floating-point NaN patterns to hold type tags and payloads.

These alternatives have two distinct performance implications.  First,
immediate representations can avoid allocating heap objects for common
small values such as small integers and, in NaN-boxed schemes, ordinary
floating-point values.  Second, even when heap objects still exist,
placing the first discriminator in the value word can avoid a memory
read merely to determine the type.  We measure how these effects
manifest on platforms in wide use today.

The experiments are organized in two stages.  Benchmark~1 separates tag
examination, payload access, and small arithmetic for low-bit tagging
and a low-tag NaN-boxing layout on two hosts.  Benchmark~2 broadens the
study using a generic box framework instantiated with heap-boxed,
low-bit-tagged, and upper-tag NaN-boxed representations across a wider
set of hosts.  This checks that the same trade-offs persist across the
variety of platforms in wide use today.

The rest of the paper is organized as follows.
Section~\ref{sec:previous-work} reviews related work on tagged
representations in Lisp, dynamic-language VMs, and computer algebra
systems.  Section~\ref{sec:representations} describes the badged-header,
low-bit-tagged, and two NaN-boxed representations measured here.
Section~\ref{sec:cost-model} gives the cost model used to interpret the
two benchmarks.  Sections~\ref{sec:benchmark1} and \ref{sec:benchmark2}
present the two benchmarks, their platforms, and their results.
Sections~\ref{sec:interpretation} and \ref{sec:caveats} discuss the
combined interpretation and limitations.

\section{Previous work}
\label{sec:previous-work}

The representation of dynamically typed values has been a systems
concern since the early Lisp implementations.  Steele's account of
PDP-10 MacLisp data representations discusses the pressure created by
limited address space, type tags, garbage collection, and the need to
represent both symbolic and numeric data efficiently
\cite{Steele1977Data}.  His companion paper on fast arithmetic in
MacLisp shows that efficient numerical computation in a Lisp system
requires close attention to representation and compiler support
\cite{Steele1977Fast}.  Gabriel's performance study of Lisp systems
later placed these concerns in a broader empirical setting, showing
that representation choices, allocation behaviour, and dispatch costs
are inseparable from the performance of symbolic programs
\cite{Gabriel1985}.  Steele's description of Common Lisp and
Queinnec's treatment of Lisp and Scheme implementation both reinforce
the same point: a uniform high-level value model must be reconciled
with non-uniform machine representations \cite{Queinnec1996,Steele1990}.

Gudeman's survey remains an important direct treatment of the
problem addressed here \cite{Gudeman1993}.  It collects a body of
implementation folklore on how dynamically typed languages represent
type information, including object headers, pointer tags, immediate
integers, and encodings based on unused floating-point
representations.  The present paper revisits this design space, but
narrows the focus to three representation strategies that continue to
occur in modern systems: header badging, low-bit pointer tagging, and
NaN-based value encodings.

The Lisp-machine and RISC-Lisp literature also studied the boundary
between software tagging and architectural support.  Steenkiste and
Hennessy compared hardware and software approaches to tags and type
checking in Lisp \cite{SteenkisteHennessy1987}.  The SPUR Lisp
architecture evaluated architectural support for efficient execution
of Lisp programs on a RISC-like processor \cite{TaylorEtAl1986}.
These studies are historically important for the present paper because
they treat type tags not as incidental encoding details, but as
operations frequent enough to influence machine design.

Object-oriented dynamic systems raised related issues: the Smalltalk-80
implementation work of Deutsch and Schiffman already showed that the
common operations in a uniform object system must be made fast enough
for practical use \cite{DeutschSchiffman1984}.  Modern dynamic-language
virtual machines inherit the same tension through immediate values,
object headers, inline caches, speculation, and de-optimization.

JavaScript virtual machines provide especially relevant recent
examples.  V8 represents values using tagged values that distinguish
small integers, or Smis, from heap object pointers.  Its pointer
compression work is a modern discussion of the trade-off between
memory footprint, tagged value format, decompression operations, and
optimization complexity \cite{SheludkoSolanes2020}.  Southern and
Renau measured the overhead of de-optimization checks in V8 and
identified both general type checks and Smi checks as frequent
low-level operations in optimized JavaScript execution
\cite{SouthernRenau2016}.  Their measurements are not about symbolic
computation, but they support the same general premise: runtime type
checks can be individually small and collectively important.

Other JavaScript engines illustrate the NaN-boxing approach.
JavaScriptCore's implementation documentation for
\texttt{JSValue} describes a NaN-encoded form that uses unused NaN
space in the IEEE 754 representation to encode non-double values
\cite{WebKitJSCJSValue}.  SpiderMonkey documentation describes
\texttt{JS::Value} as the representation for JavaScript values, and
older internals documentation describes platform-specific NaN-boxing
formats, including nunboxing and punboxing
\cite{MozillaSpiderMonkeyDocs,SpiderMonkeyInternals}.  These are
implementation documents rather than conventional papers, but they
are directly relevant because the exact encoding choices are
implementation-level facts.

The architecture literature has also treated dynamic type checking as
frequent enough to merit hardware support.  Anderson, Fortuna, Ceze,
and Eggers proposed checked-load instructions for JavaScript
type-checking in the Nitro JavaScript JIT, motivated by the dynamic
instruction cost of software type checks \cite{AndersonEtAl2011}.
More recent work by Melan{\c{c}}on, Serrano, and Feeley explicitly
frames the current design space in terms of tagged pointers,
NaN-boxing, and related encodings, and proposes float self-tagging for
dynamically typed languages \cite{MelanconSerranoFeeley2025}.

Computer algebra systems add their own design pressures.  The early design
of Maple emphasized compactness, portability, and efficient internal
data structures as central design criteria \cite{CharEtAl1983}.  A
later paper on the design and performance of Maple identified the
relationship between kernel operations, interpreted library code, and
data representation as a practical systems issue \cite{CharEtAl1984}.
More recent work on Maple's sum-of-products and \textsc{poly} data
structures gives a detailed account of how mathematical objects are
represented and why specialized structures matter for performance
\cite{MonaganPearce2014}.  Standard computer algebra texts naturally
concentrate on algorithms \cite{GeddesCzaporLabahn1992}, but the
implementation of those algorithms depends heavily on the cost of the
primitive operations used to represent, classify, and traverse
objects.

Maple also provides a direct computer-algebra example of the immediate
scalar trade-off considered here. Its implementation has had to
support both object traversal and high-frequency arithmetic inside a
large, long-lived system. Implementation-history information supplied
privately by Maple kernel developers indicates that immediate integers
using low bit tagging were introduced in Maple 6. More recent Maple versions have extended
the immediate-value design to include selected floating-point values.
In Maple 2025, a range of immediate floating-point values were added using 
low tag bits at the cost of reducing the
immediate integer range.
This is exactly the kind of trade-off examined here: some immediate
range is sacrificed, but common scalar values can avoid allocation and
can often be classified without opening a heap object.

This paper lies between dynamic-language VM implementation
and symbolic computation implementation.  From the VM literature, it
takes the question of how to encode type information efficiently in a
machine word or an object header.  From computer algebra, it takes
the workload assumption that shallow type tests and small arithmetic
operations occur frequently inside symbolic kernels.  The contribution
is a controlled re-measurement of a familiar trade-off: whether it is
better to read a badge from an object header or to pay the
bit-manipulation cost of a tagged value.

\section{Representations Tested}
\label{sec:representations}

This section describes the three representation strategies implemented
in the benchmark.  The benchmark is deliberately small and does not
attempt to reproduce the full object system of a production language
runtime.  Its purpose is to isolate the first-order cost of
classification of objects and small payload operations.

\subsection{Badged object headers}

In the badged-header representation, every value is represented by a
pointer to a heap object.  The first word of the object is a header.
In the benchmark, the header contains a small type code, or badge,
identifying the object as a natural number, an integer, a double-precision floating-point number, or a cons cell.
The remaining word or words contain the payload.

In the following we use the \texttt{C/C++} terminology for machine type names.
A boxed natural number or integer contains a badge and a
\texttt{uint64\_t} or \texttt{int64\_t}
payload.  A boxed double contains a double badge and a
double-precision payload.  A cons cell contains a cons badge and two
generic fields. This is shown in Figure~\ref{fig:header-layout}. Each object is aligned on a machine-word
boundary.  Real systems have many more object types, but this small set
captures the essential features needed in the micro-benchmark: scalar
payloads, floating-point values, and compound references.

The representation is simple and general.  It is also close in spirit
to managed runtimes in which object headers contain type descriptors,
sizes, forwarding state, locking state, or garbage-collection metadata
\cite{JonesHoskingMoss2011}.  Its cost for this experiment is that a
pure type test must follow the pointer and read the first word of the
object.

\input{figures/header-layout}

\subsection{Low-bit pointer tagging}

The low-bit representation uses the fact that heap objects are aligned.
On now-pervasive byte-addressed architectures, heap objects occur at
addresses where at least the three low-order bits of object pointers are
zero.  These bits can therefore be used to encode small type tags.

Integral values are encoded by up-shifting them by three bits.   This reduces the range of integer types, but supports the important subset of small values.   Production systems usually have an arbitrary precision big integer type, so this loss of bits for small values means that some word-size integers need to be represented as heap-allocated big integers.  Note that arithmetic need not fully decode the arguments.   For example, if values do not overflow and the low $s$ bits are used for tagging, we have
\begin{align*}
    \encode(a + b) &= 2^{s}(a + b) + \mathrm{tag} \\
                   &= (2^{s} a + \mathrm{tag})
                    + (2^{s} b + \mathrm{tag})
                    - \mathrm{tag} \\
                  &=\encode a + \encode b - \mathrm{tag}
\end{align*}

\input{figures/lowtag-layout}

Low-order type codes are not the only possibility.
High-order bit type codes have also been used, and remain interesting
when an implementation has control of real memory or can arrange
segment-like mappings in virtual memory.  As one historical example,
muLisp used segment registers so that cars and cdrs could be accessed
from the same address through different segments
\cite{StoutemyerPrivate}.  The present benchmark does not attempt to
measure such schemes, but they are a similar representation.

\subsection{NaN-boxing}
\label{subsec:nanboxing}

NaN-boxing uses the structure of IEEE double-precision values, exploiting the range of ``not-a-number'' values defined by the IEEE 754 standard~\cite{Goldberg1991,IEEE7542019}.  A runtime can reserve a
subset of these patterns for non-floating-point values.  Ordinary doubles can
then remain ordinary double-precision bit patterns, while integers,
pointers, and other immediate data values occupy selected quiet-NaN payloads.

We consider two NaN-boxing layouts, shown together in
Figure~\ref{fig:nanbox-layouts}.  The low-tag layout puts the type tag
in the low-order bits and shifts non-floating payloads to make room for
it.  This layout is close to low-bit pointer tagging and is useful for
comparing the extra cost of NaN classification with the familiar
low-bit-tag operations.  
This is the layout used in our first benchmark.

The upper-tag layout puts the type tag
in bits 50--48 and keeps the lower 48 bits as an unshifted payload.
This layout is usually more convenient when 48-bit canonical pointers (with all top bits replicating bit 47)
or 48-bit immediate signed integers are common: the payload can be recovered by a simple upshift followed by an arithmetic downshift.
It is the layout used in the later multi-host comparison.

\input{figures/nanbox-layouts}

A 48-bit payload is sufficient for many present-day 64-bit address
conventions.  On x86-64 systems using the traditional four-level page
table regime, virtual addresses are canonical: bits 63--48 replicate
bit 47, so the effective address payload is 48 bits.  AArch64 systems
commonly use 48-bit virtual-address configurations, although exact
address-size and top-bit conventions depend on the operating system
and hardware configuration.  RISC-V virtual-memory schemes such as
Sv39 and Sv48 likewise use sign-extension from the highest implemented
virtual-address bit.  Thus, on many current systems, a 48-bit payload
is enough to reconstruct a canonical pointer.  Larger virtual address
spaces do not necessarily invalidate the layout.  For example, x86-64
with five-level page tables extends canonical virtual addresses to
57 bits, but heap-object alignment can be used to recover some of
those bits: if objects are aligned to 8 or 16 bytes, the low-order
zero bits need not be stored and can be restored by shifting during
decoding.

\section{A simple cost model}
\label{sec:cost-model}

The experiments are interpreted using a deliberately simple cost model.
The purpose of the model is not to predict exact timings, but to
separate the effects that the benchmark is intended to expose.

For a header-badged object, a type test has the approximate form
\[
T_{\mathrm{header}}
  \approx
  T_{\mathrm{load\ ref}}
  + T_{\mathrm{load\ header}}
  + T_{\mathrm{classify}}
  + T_{\mathrm{branch}} .
\]
The important term is the header load.  It is dependent on the value
loaded from the array.  If the referenced object is not in a nearby
cache level, the type test may be limited by memory latency rather
than by the small number of instructions used to inspect the badge.

For a low-bit tagged value, the corresponding cost is roughly
\[
T_{\mathrm{lowtag}}
  \approx
  T_{\mathrm{load\ word}}
  + T_{\mathrm{mask/shift}}
  + T_{\mathrm{classify}}
  + T_{\mathrm{branch}} .
\]
The additional operations are local integer operations on an already
loaded word.  They may increase instruction count, but they avoid the
dependent object-header load.
Note that to use an un-shifted low-bit tagged reference, there is no need to mask off the lower bits --- the misalignment of the lower bits can be subtracted at compile time from the field offsets.

For a NaN-boxed value, the cost can be summarized as
\[
T_{\mathrm{nanbox}}
  \approx
  T_{\mathrm{load\ word}}
  + T_{\mathrm{nan\ classify}}
  + T_{\mathrm{extract}}
  + T_{\mathrm{branch}} .
\]
The key question is whether this local classification and extraction
cost is smaller than the memory-probe penalty of opening a boxed
object.

The model suggests three expectations.  First, when the working set is
large and object order is randomized, header badging should suffer
from the dependent memory access.  Second, when type discrimination is
followed by little arithmetic, low-bit tagging and NaN-boxing should
benefit from avoiding the header load.  Third, when small arithmetic
does follow the type test, immediate representations should retain an
advantage if the payload itself is in the value word.

The generated code for the primitive accessors gives a more concrete
view of these costs, as seen in
Sections~\ref{sec:benchmark1} and~\ref{sec:benchmark2}.

\section{Benchmark 1: Low-bit tags and low-tag NaN boxing}
\label{sec:benchmark1}

The first benchmark, described in this section, compares low-bit
tags, a low-tag NaN layout and reading header fields of values stored in ordinary \texttt{C++} library vectors.  
It tests tag examination, payload access, and simple
arithmetic on one x86-64 and one AArch64 platform.

\subsection{Design}

The first benchmark constructs a large base array of heap objects.  Each
object is one of three kinds: a boxed integer, a boxed double, or a
boxed cons cell.  Object kinds are selected pseudorandomly with equal
probability.  The benchmark then shuffles the array to reduce artificial
locality.  The shuffle preserves the relative order of doubles so that
the floating-point sum is deterministic across runs.

After the base array has been constructed, the benchmark constructs two
additional word arrays.  The first is a low-bit tagged array.  In some
passes it contains tagged pointers; in the immediate-integer pass it
contains immediate integers for integer values and tagged pointers for
other values.  The second is a low-tag NaN-boxed array containing
ordinary double bit patterns for doubles, NaN-boxed immediate integers
for integer values, and NaN-boxed pointers for cons cells.  This
benchmark uses standard library vectors and does not use templates to
abstract over the representation.

\input{tables/phases.tex}
\input{tables/primitive-instruction-counts.tex}
The benchmark uses deterministic payload generators.  Integer payloads
cycle through the values 1 to 10.  Cons-cell fields cycle through small
integer values.  Double values begin at 1.0 and are repeatedly
multiplied by 1.0000001.  The exact numerical values are not important
for the type-test comparison; they make the sums observable and
repeatable.

For all reported sizes, the generated populations are close to equal
thirds.  For the largest run, with $10^9$ values, the counts were
333,325,192 cons cells, 333,349,758 integers, and 333,325,050 doubles.

Table~\ref{tab:phases} gives the eight timed passes for the first
benchmark.  The distinction between P3 and P5 is particularly important.
P3 uses low-bit tags to classify pointers but still dereferences integer
objects to obtain the payload.  P5 stores integer payloads immediately
in the tagged word.  The benchmark was run at array sizes $10^n,
n \in 3..9$, with five repetitions at each size.  The reported phase
times are the means of the five repetitions.

The instruction counts for the key representation operations are shown in
Table~\ref{tab:primitive-instruction-counts}.  These counts are taken from the generated assembly for the stand-alone accessors and from the hot traversal loops.

The header representation
is not instruction-heavy in the narrow sense.  Once a heap object has
been allocated, installing the tag and payload requires only one or two
store instructions, and reading a tag from a known object pointer is a
single byte load.  The important difference is the memory dependence in
the traversal loop: reading a header tag from an array element requires
first loading the object pointer and then performing a dependent load
from the heap object.  Low-bit tagging and NaN-boxing replace this
dependent heap load by local integer operations on the value word.  The
exact instruction sequence depends on the NaN-boxing layout, but the
common point is that ordinary doubles must be distinguished from boxed
non-doubles before extracting the tag or payload.  That classification
costs several local instructions, but it still avoids opening a heap
object merely to discover the type.

\subsection{Platforms}

Benchmark~1 was run on two large-memory platforms, chosen to
provide one AArch64 host and one x86-64 host.
These two machines are summarized in
Table~\ref{tab:benchmark1-platforms}.
The table shows the number of cores / threads even though the experiment is single threaded.
\input{tables/benchmark1-platforms.tex}

\subsection{Results and analysis}

Tests were performed with arrays of values with $10^i$ elements, $i \in 3..9$.
The smallest runs are useful mainly as sanity checks, though their absolute
times are too small for stable interpretation.
The main conclusions are drawn from
$10^6$ through $10^9$, and especially from $10^7$ through $10^9$.

Tables~\ref{tab:aarch64-cpp-ns-results} and
\ref{tab:x86-cpp-ns-results} give the mean time per value for the four
largest sizes on the two Benchmark~1 hosts: grice (Apple M3 Max,
AArch64) and cswaterloo (AMD EPYC 9554 under KVM/QEMU, x86-64).  Times
are in nanoseconds per value, computed from the mean of five
repetitions.
The five repetitions are sufficiently stable for the large AArch64 runs:
for $n \geq 10^7$, most coefficients of variation are below five
percent.  The x86-64 runs show larger absolute variation in some
heap-reading passes, consistent with virtualization and memory-latency
effects, but the qualitative conclusions do not depend on a single
outlying run.
\input{tables/aarch64-cpp-ns-results.tex}
\input{tables/x86-cpp-ns-results.tex}
\input{tables/cross-platform-ratios.tex}

The first benchmark shows the main pattern on both platforms.  At
$10^9$ values, header-based type counting is much slower than
classification from the value word: P1/P2 is 14.5 on AArch64 and 36.8
on the x86-64 platform.  

Header-based integer summation is also much
slower than immediate integer summation, with P4/P5 equal to 6.3 and 7.5 for low-bit type codes and 8.9 and 4.3 for low-tag NaN-boxed values on AArch64 and x86-64 respectively.  

The low-tag NaN-boxed integer summation
pass P6 is somewhat more expensive than the low-bit immediate pass on
x86-64, but it is still far faster than reading integer payloads through
header badges.

The comparison between P4 and P3 shows the difference between obtaining tags from the value word versus the object header.  Both passes ultimately
read integer payloads from heap objects, but P3 identifies integer
values from low-bit tags while P4 reads object headers to identify them.
At $10^9$ values, P4/P3 is 2.52 for AArch64 and 1.59 for x86-64.  Thus,
even when selected integer payloads still reside in heap objects, using
the object header as the first discriminator remains significantly more
expensive than using the value word.

\section{Benchmark 2: Generic box representations across hosts}
\label{sec:benchmark2}

The second benchmark compares generic box representations across a wider
set of hosts.  The NaN-boxed case uses the upper-tag 48-bit-payload layout
of Figure~\ref{fig:nanbox-layouts}; ordinary doubles remain ordinary IEEE
values, and upper NaN payload bits carry a type code for non-double values.
This is the broader generality study, testing whether the same
representation trade-offs persist across different memory hierarchies and
deployment classes.
The benchmark source code is archived as
\cite{GenericBoxTest1}.

\subsection{Design}

The benchmark is written against a \texttt{C++} \texttt{Box} ``concept'' and is
instantiated by three representations.  
\begin{itemize}
\item 
\textsc{HaBox} is the heap
baseline: \texttt{Nat}, \texttt{Int}, \texttt{Flo}, and cons cells are
all heap-allocated and carry header tags.  
\item
\textsc{LoBox} uses low-order
type codes: \texttt{Nat} and \texttt{Int} are immediate values, conses
are low-bit tagged pointers, and \texttt{Flo} remains heap boxed.
\item
\textsc{NanBox} uses the upper-tag NaN-boxing layout: \texttt{Nat} and
\texttt{Int} are immediate 48-bit payloads, \texttt{Flo} is an ordinary
unboxed double unless it is a canonicalized NaN, and conses are boxed
references encoded in the NaN payload.
\end{itemize}

Each benchmark run constructs an array of 
values rather than using a standard library vector.  This difference is not essential to the representation comparison.
The array contains an equal number of unsigned integer, signed integer, floating-point and cons values.  
The cons cells contain two integers.
After the array is allocated, the order of the entries is randomized to avoid predictable memory access patterns.

The benchmark
measures three loops: \texttt{tags}, which
only counts value tags; \texttt{grouped~+}, which adds the signed and unsigned integer 
values using unboxed C accumulators and groups 25 array accesses in each loop iteration; and \texttt{boxed~+}, which
performs the same additions while keeping the running sums as boxed
values and accessing only one slot per iteration.  
This latter loop is intended to model generic arithmetic more
closely, since the intermediate results remain dynamically checkable and occur within other code.

\subsection{Platforms}

The second benchmark was run on a wider set of hosts.  These platforms
differ not only in instruction set, but also in memory hierarchy,
operating system, compiler, virtualization status, and available
address-space features.  The comparison should therefore be read as a
comparison of disparate modern platforms rather than as a pure ISA
comparison.  Table~\ref{tab:platform-rationale} gives the reason each platform is
included in the comparison; Table~\ref{tab:platforms} summarizes the host
properties used to interpret the measurements.

The x86-64 server platform reported 52-bit physical addresses and
57-bit virtual addresses, with the \texttt{la57} feature present.  Thus
it is already an example of the larger-address-space situation discussed
in Section~\ref{subsec:nanboxing}: a 48-bit NaN-box payload is no longer enough to store every
canonical virtual address directly.  For aligned heap objects, however,
compact pointer encodings remain possible by omitting low-order zero
bits and restoring them during decoding.
\input{tables/platform-rationale.tex}
\input{tables/platforms.tex}

\subsection{Results and analysis}

Table~\ref{tab:benchmark2-o3-crosshost} gives the main cross-host
comparison for Benchmark 2 at optimization level \texttt{-O3} and $10^8$ values.  Each
entry is a speedup ratio relative to \textsc{HaBox}, so larger values
mean that the heap-header representation is slower than either the
low-bit or NaN-boxed representation.  The abbreviations H/Lo and H/Nan
denote \textsc{HaBox}/\textsc{LoBox} and
\textsc{HaBox}/\textsc{NanBox}, respectively.

\input{tables/benchmark2-o3-crosshost.tex}

The cross-host table is intentionally more compact than the full log
matrix.  It uses one canonical size and optimization level, so additional
processors can be added as rows without re-engineering the presentation.
This benchmark should be read as the broader generality study following
the more detailed two-host mechanism study of Benchmark~1.
Table~\ref{tab:benchmark2-summary-ranges} summarizes the same data as
ranges.  This makes the main pattern visible without asking the reader
to compare every machine manually.

\input{tables/benchmark2-summary-ranges.tex}

Across the measured platforms, the ratios fall into three broad regimes.
For tag-only loops, heap-header classification is about 10--32 times
slower than low-bit classification on the main contemporary hosts, and
about 6--19 times slower than NaN-boxed classification.  Including the
older Core i3 and Core i7 widens these ranges, as one
would expect for older memory-system observations.  For
\texttt{grouped +}, where payload access and arithmetic are included,
the ratios are much smaller, typically around two to four.  For
\texttt{boxed +}, where intermediate results remain dynamically
checkable boxed values, the heap representation is again several times
slower.

Table~\ref{tab:nanbox-lobox-ratios-o3-1e8} isolates the cost of
NaN-boxing relative to low-bit tagging for the same canonical
\texttt{-O3}, $10^8$ comparison.  Low-bit tagging is generally cheaper
for pure tag examination.  In the arithmetic loops, however, the gap is
much smaller, and NaN-boxing has the separate advantage that ordinary
doubles need not be heap allocated.

\input{tables/nan_lobox_ratios_O3_1e8.tex}

The generated code gives a useful explanation of these timing results.
Tables~\ref{tab:alt-form-instruction-counts-o3} and
\ref{tab:alt-access-instruction-counts-o3} give static fast-path
instruction counts for forming values and for tag/payload access in the
\texttt{-O3} assembly.  These are instruction counts, not cycle counts.
They do not model branch prediction, cache misses, allocator costs,
micro-op decomposition, or instruction latency.  They are included to
explain which operations are local bit manipulations and which
operations require heap access.

The instruction counts reinforce the timing interpretation.  Low-bit tag
extraction is the simplest classification operation.  NaN-boxing
requires additional local work to distinguish ordinary doubles from
boxed non-doubles and to extract the tag field, but its payload
extractions are comparable once the value has been classified.
Header-badged objects are not expensive because the visible instruction
sequence is long; they are expensive because the tag is obtained by a
dependent memory read from an object that may not be nearby in cache.

\input{tables/alt_form_instruction_counts_O3.tex}
\input{tables/alt_access_instruction_counts_O3.tex}

\section{Interpretation}
\label{sec:interpretation}

The results support an expected conclusion: for shallow tests
over large value streams, the first discriminator should preferably
be in the value word itself.  A header badge is flexible and simple,
but it requires a dependent load from the heap object before the
system even knows what kind of value it has.  In this benchmark, that
extra dependency is much more expensive than the masking and shifting
needed for low-bit tags or the classification and extraction needed
for NaN-boxed values.

There are two separable benefits of encoding tags and small values.  The first is avoiding heap
allocation for common values.  Immediate integers and unboxed doubles
reduce storage traffic before traversal even begins, and they reduce
the number of objects later visited by the memory system.  The second
benefit is avoiding a memory read during classification.  Even when a
heap object still exists, as in the tagged-pointer integer-sum pass,
putting the first type discriminator in the value word avoids using
that heap object as the type oracle.  The experiments show both effects: P3 lies between P4 and P5, while
P5 and the alternate NaN-boxing measurements show the benefit of
keeping frequent numeric values out of separate heap objects.

The upper-tag NaN-boxing experiment adds a further comparison.  The
48-bit-payload layout 
keeps tag extraction and payload extraction simple.  With the
loop overhead reduced, its large-array tag-counting cost is close to
that of low-bit tagging.

For symbolic computation, this matters because many operations are
not dominated by heavy arithmetic.  Expression traversal, testing for
small integer coefficients or exponents, distinguishing atoms from compound
expressions, and dispatching among shallow object forms may perform
many type tests followed by little work.  In that setting, the type
test is part of the inner loop.

The low-bit versus NaN-boxing comparison should also be read against
changes in setting.  Traditional Lisp-like symbolic applications
often perform little double-precision arithmetic in their generic value
paths, so low-order type codes remain a very good choice.  They give
the cheapest tag examination and simple immediate integers.  In
contrast, virtual machines such as those for JavaScript and Python
frequently carry dynamically typed numeric values across generic
interfaces; JavaScript numbers are normally double-precision values,
and Python floats are also represented as doubles.  For such workloads,
NaN-boxing is not much more expensive in the arithmetic loops measured
here, and it avoids heap-boxing ordinary doubles.  In a generic value
array, a NaN-boxed double occupies one 8-byte word rather than an
8-byte pointer plus a 16-byte heap object, exactly one third the space
before allocator overhead is counted.

The benchmark also suggests why hybrid representations remain
attractive.  Small integers are strong candidates for immediate
representation.  Double-precision values may benefit from direct or
NaN-boxed representation when they occur frequently.  Larger compound
objects can remain boxed and carry headers, since their payload will
often have to be accessed anyway.  The main design question is where
to place the boundary between values that should be classified from
the value word and values whose type information can safely remain in
the object header.

\section{Caveats}
\label{sec:caveats}

These tests are micro-benchmarks.  They isolate representation costs
but do not reproduce the full behaviour of software systems.
Real systems allocate, garbage collect, and operate on a wide variety of data structures.
The results therefore identify mechanisms
rather than predict whole-system speedups.

The first benchmark keeps several alternative representations live at
once.  This is useful for comparing passes under similar generated
data, but it means that the measured process memory footprint is not
the memory footprint of any one representation.

The value distributions are synthetic.  Different symbolic workloads
will contain very different proportions of small integers,
floating-point numbers, symbols, lists, polynomials, matrices, and
other objects.  The equal proportion mixtures used here are to surface measurable effects.

Compiler optimization is another concern.  Changes in source
code can alter instruction selection, branch layout, register
allocation, and vectorization.  
Generated code should still be
inspected when interpreting surprising results.  
This is especially
important because the benchmark is small enough that an optimizing
compiler may transform code substantially.

Cache state is also relevant.  Some representation arrays are
materialized immediately before the corresponding timed pass.  For
very large arrays this does not keep the full array hot in cache, but
it may still influence the state of nearby cache levels.  A more
elaborate version of the experiment could run each pass in a separate
process or randomize pass order.  The present measurements are best
understood as steady traversals of large materialized arrays.

Some of the x86-64 measurements were obtained in non-native or
virtualized environments.  The KVM/QEMU host reports 64 virtual CPUs as
64 sockets with one core each, and the panamint i7 result was obtained
under Cygwin on Windows.  The benchmark itself is single-threaded, but
virtualization, operating-environment effects, scheduling, and the
virtualized memory hierarchy may affect absolute timings.  The
comparison with other platforms should therefore be used primarily for
qualitative cross-checking of the representation trade-off, not for
ranking the processors.

Finally, the results are platform-specific.  The relative costs of
integer operations, dependent loads, branch mispredictions, and memory
latency differ across processors.  The conclusions should therefore
be read only as evidence about the measured contemporary platforms.

\section{Conclusion}
\label{sec:conclusion}

The experiments here compare badged heap objects, low-bit pointer tags,
and two NaN-boxing layouts under a constructed workload: type tests,
small integer and floating-point access, cons traversal, and boxed
intermediate arithmetic.  The main lesson is that memory accesses to an
unexamined object remain expensive relative to local bit-manipulation
operations.  That is, it is still worthwhile to have examinable tags to avoid memory references, and to ``look before you leap.''

Our results show that different representations will be preferred for different settings.  Low-bit
type codes remain an excellent default for mostly symbolic Lisp-like and
computer-algebra workloads in which small integers, symbols, and
container nodes dominate, and generic double-precision arithmetic is not
central.  NaN-boxing becomes attractive when significant floating-point
arithmetic is expected, since it is close in access cost to low-bit
tagging while being significantly more space efficient for ordinary
double-precision values.

The compiler characteristics also matter.  Compiled code for a dynamic-language
or symbolic-computation VM can often keep heavily used quantities
unboxed inside specialized functions and box only at generic interfaces
or de-optimization points.  

What we have seen in our experiments is broadly along the lines we
expected, and now we can be quantitative.  Across a broad range of
platforms in use today:
If only a tag check is needed, tagged boxing schemes can be about
10--30 times faster than examining headers, depending on the details.
Working with immediate values where possible not only saves the
cost of memory management, using them is about 2--5 times faster than working
with stored payloads.
Thus expending effort on boxing schemes remains worthwhile
today.  These trade-offs will continue to change, so they should be revisited again from time to time.

\section*{Acknowledgements}
The author thanks Arthur Norman for providing the Raspberry Pi and Intel Core i7 timing
data used in the multi-host comparison.  Log file analysis and some parts of text preparation were done with directed use of ChatGPT.
{\small
\bibliographystyle{splncs04}
\bibliography{main}
}
\end{document}

%% file: figures/header-layout.tex
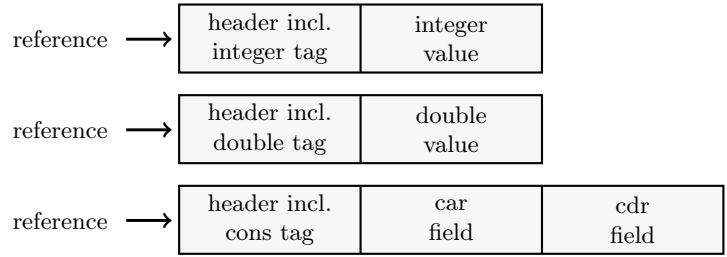
\begin{figure}[!t]
\begin{center}
\begin{tikzpicture}[x=1cm,y=0.88cm,every node/.style={font=\footnotesize}]
  \newcommand{\twoword}[4]{%
    \node[anchor=east] at (0.65,#1+0.50) {#2};
    \begin{scope}[shift={(1.48,#1)}]
      \fill[black!3] (0,0) rectangle (4.80,1.02);
      \draw[thick] (0,0) rectangle (4.80,1.02);
      \draw[thick] (2.40,0) -- (2.40,1.02);
      \node[align=center] at (1.20,0.51) {#3};
      \node[align=center] at (3.60,0.51) {#4};
    \end{scope}
    \draw[line width=1.1pt,->] (0.78,#1+0.50) -- (1.38,#1+0.50);
  }
  \newcommand{\threeword}[5]{%
    \node[anchor=east] at (0.65,#1+0.50) {#2};
    \begin{scope}[shift={(1.48,#1)}]
      \fill[black!3] (0,0) rectangle (7.20,1.02);
      \draw[thick] (0,0) rectangle (7.20,1.02);
      \draw[thick] (2.40,0) -- (2.40,1.02);
      \draw[thick] (4.80,0) -- (4.80,1.02);
      \node[align=center] at (1.20,0.51) {#3};
      \node[align=center] at (3.60,0.51) {#4};
      \node[align=center] at (6.00,0.51) {#5};
    \end{scope}
    \draw[line width=1.1pt,->] (0.78,#1+0.50) -- (1.38,#1+0.50);
  }
  \twoword{2.72}{reference}{header incl.\\integer tag}{integer\\value}
  \twoword{1.36}{reference}{header incl.\\double tag}{double\\value}
  \threeword{0.00}{reference}{header incl.\\cons tag}{car\\field}{cdr\\field}
\end{tikzpicture}
\end{center}
\caption{Badged object headers.  The type test follows the reference
and reads the first word of the object.  Integer and double boxes use
one header word and one value word; a cons cell uses one header word
and two field words.\spacetune{\newline}}
\label{fig:header-layout}
\end{figure}

%% file: figures/lowtag-layout.tex
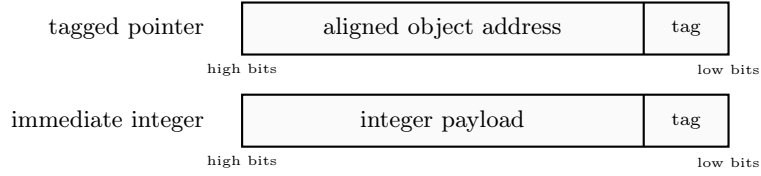
\begin{figure}[!t]
\centering
\begin{tikzpicture}[x=1.10cm,y=0.78cm,every node/.style={font=\small}]
  \def\h{0.88}
  \node[anchor=east] at (0,2.00) {tagged pointer};
  \draw[thick, fill=black!2] (0.35,1.56) rectangle (6.20,1.56+\h);
  \draw[thick] (5.18,1.56) -- (5.18,1.56+\h);
  \node[align=center] at (2.76,2.00) {aligned object address};
  \node[align=center,font=\scriptsize] at (5.69,2.00) {tag};
  \node[font=\tiny] at (0.35,1.30) {high bits};
  \node[font=\tiny] at (6.20,1.30) {low bits};

  \node[anchor=east] at (0,0.44) {immediate integer};
  \draw[thick, fill=black!2] (0.35,0) rectangle (6.20,\h);
  \draw[thick] (5.18,0) -- (5.18,\h);
  \node[align=center] at (2.76,0.44) {integer payload};
  \node[align=center,font=\scriptsize] at (5.69,0.44) {tag};
  \node[font=\tiny] at (0.35,-0.26) {high bits};
  \node[font=\tiny] at (6.20,-0.26) {low bits};
\end{tikzpicture}
\caption{Low-bit tagging.  For a tagged pointer, the low-order
tag identifies the kind of value reached through the pointer.  For an
immediate integer, the shifted value is stored directly in the word;
the tag can also distinguish signed and unsigned integer encodings.}
\label{fig:lowtag-layout}
\end{figure}

%% file: figures/nanbox-layouts.tex
\begin{figure}[!t]
\centering
\resizebox{0.98\linewidth}{!}{%
\begin{tikzpicture}[x=1cm,y=1cm,every node/.style={font=\small}]
  \def\w{10.8}
  \def\hh{0.98}
  \node[anchor=east] at (-0.32,3.55) {ordinary double:};
  \draw[thick, fill=black!3] (0,3.06) rectangle (\w,3.06+\hh);
  \node at (5.4,3.55) {bits 63--0: IEEE double-precision value};

  \node[anchor=east, align=right] at (-0.32,1.92)
    {low-tag NaN-boxed\\non-double:};
  \draw[thick, fill=black!8] (0,1.43) rectangle (0.55,1.43+\hh);
  \draw[thick, fill=black!5] (0.55,1.43) rectangle (3.10,1.43+\hh);
  \draw[thick, fill=black!2] (3.10,1.43) rectangle (9.35,1.43+\hh);
  \draw[thick, fill=black!12] (9.35,1.43) rectangle (\w,1.43+\hh);
  \node[rotate=90, font=\scriptsize] at (0.275,1.92) {sign};
  \node[align=center, font=\scriptsize] at (1.825,1.92)
    {quiet-NaN\\discriminator};
  \node[align=center] at (6.22,1.92) {shifted value or pointer};
  \node[align=center, font=\scriptsize] at (10.075,1.92) {low-bit\\tag};
  \node[font=\tiny] at (0,1.17) {63};
  \node[font=\tiny] at (0.55,1.17) {62};
  \node[font=\tiny] at (3.10,1.17) {51};
  \node[font=\tiny] at (9.35,1.17) {2};
  \node[font=\tiny] at (\w,1.17) {0};

  \node[anchor=east, align=right] at (-0.32,0.49)
    {upper-tag NaN-boxed\\non-double:};
  \draw[thick, fill=black!8] (0,0) rectangle (0.55,\hh);
  \draw[thick, fill=black!5] (0.55,0) rectangle (3.10,\hh);
  \draw[thick, fill=black!12] (3.10,0) rectangle (4.35,\hh);
  \draw[thick, fill=black!2] (4.35,0) rectangle (\w,\hh);
  \node[rotate=90, font=\scriptsize] at (0.275,0.49) {sign};
  \node[align=center, font=\scriptsize] at (1.825,0.49)
    {quiet-NaN\\discriminator};
  \node[align=center, font=\scriptsize] at (3.725,0.49)
    {tag};
  \node[align=center] at (7.575,0.49)
    {48-bit value\\\scriptsize pointer or immediate};
  \node[font=\tiny] at (0,-0.26) {63};
  \node[font=\tiny] at (0.55,-0.26) {62};
  \node[font=\tiny] at (3.10,-0.26) {51};
  \node[font=\tiny] at (4.35,-0.26) {48};
  \node[font=\tiny] at (\w,-0.26) {0};
\end{tikzpicture}}
\caption{Two NaN-boxing layouts.  Both preserve ordinary IEEE
 double-precision values.  The low-tag layout stores the tag in the
low-order bits of the NaN-boxed non-double value and shifts payloads to
make room.  The upper-tag layout stores the type tag in bits 50--48 and
uses the lower 48 bits for a pointer or immediate value.}
\label{fig:nanbox-layouts}
\end{figure}
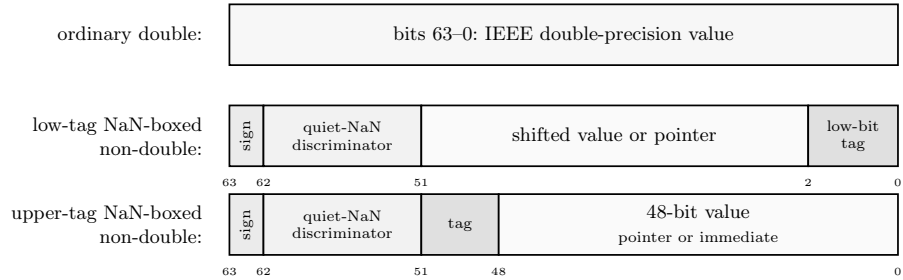

%% file: tables/phases.tex
\begin{table}[t]
\centering
\caption{The timed passes used in Benchmark~1.  P3 and P5 separate two
 effects: P3 classifies integers from the value word but still reads
 their payloads from heap objects, while P5 stores integer payloads
 immediately in the value word.}
\label{tab:phases}
\footnotesize
\setlength{\tabcolsep}{3pt}
\begin{tabular}{lll}
\toprule
Pass & Operation & Representation and payload location\\
\midrule
P1 & Count conses/integers/doubles
   & Header-badged heap objects; \\
   && tag in object header\\
P2 & Count conses/integers/doubles
   & Low-bit tagged words; \\
   &&  tag in value word\\
P3 & Sum integer payloads
   & Low-bit tagged pointers; \\
   &&  tag in value word, payload in heap\\
P4 & Sum integer payloads
   & Header-badged heap objects; \\
   &&  tag and payload read from heap\\
P5 & Sum integer payloads
   & Low-bit immediate integers; \\
   &&  tag and payload in value word\\
P6 & Sum integer payloads
   & Low-tag NaN-boxed immediate integers; \\
   &&  tag and payload in value word\\
P7 & Sum double payloads
   & NaN-boxed ordinary double values\\
P8 & Sum integers and doubles
   & NaN-boxed mixed numeric values\\
\bottomrule
\end{tabular}
\end{table}

%% file: tables/primitive-instruction-counts.tex
\begin{table}[!t]
\centering
\footnotesize
\setlength{\tabcolsep}{2pt}
\caption{AArch64 instruction counts for 
operations used in Benchmark 1.}
\label{tab:primitive-instruction-counts}
\begin{tabular}{llll}
\toprule
Representation & Operation & Representative instructions & Count\\
\midrule
Header object
  & construct after allocation
  & \texttt{strb} tag; store payload
  & 1--2 stores\\
Header object
  & read tag from array
  & \texttt{ldr} pointer; \texttt{ldrb} tag
  & 2\\
Header object
  & read tag, pointer known
  & \texttt{ldrb} tag
  & 1\\
Header object
  & extract integer payload
  & \texttt{ldr} payload
  & 1\\
Header object
  & extract double payload
  & \texttt{ldr} payload
  & 1\\
\midrule
Low-bit tag
  & tag pointer, generic
  & \texttt{uxtw}; \texttt{and}; \texttt{orr}
  & 3\\
Low-bit tag
  & tag pointer, fixed tag
  & \texttt{and}; \texttt{orr}
  & 2\\
Low-bit tag
  & tag unsigned integer
  & \texttt{lsl}; \texttt{orr}
  & 2\\
Low-bit tag
  & read tag, word known
  & \texttt{and}
  & 1\\
Low-bit tag
  & read tag from array
  & \texttt{ldr} word; \texttt{and}
  & 2\\
Low-bit tag
  & extract pointer payload
  & \texttt{and}
  & 1\\
Low-bit tag
  & extract integer payload
  & \texttt{lsr}
  & 1\\
\midrule
NaN-box
  & box pointer
  & \texttt{and}; \texttt{orr}; \texttt{orr}
  & 3\\
NaN-box
  & box unsigned integer
  & \texttt{orr}; \texttt{orr}
  & 2\\
NaN-box
  & box double
  & \texttt{fmov}
  & 1\\
NaN-box
  & read type, known boxed
  & \texttt{ubfx}
  & 1\\
NaN-box
  & read type from array, known boxed
  & \texttt{ldr} word; \texttt{ubfx}
  & 2\\
NaN-box
  & full classify, word known
  & \texttt{and}; \texttt{ubfx}; \texttt{cmp}; branch/select
  & 4--5\\
NaN-box
  & full classify from array
  & \texttt{ldr}; \texttt{ubfx}; \texttt{and}; \texttt{cmp}; branch
  & 5\\
NaN-box
  & extract unsigned payload
  & \texttt{and}
  & 1\\
NaN-box
  & extract pointer payload
  & \texttt{sbfx}
  & 1\\
NaN-box
  & extract double payload
  & \texttt{fmov}
  & 1\\
\bottomrule
\end{tabular}
\end{table}

%% file: tables/benchmark1-platforms.tex
\begin{table}[!t]
\centering
\caption{Benchmark~1 platforms for
pass-level measurements of P1--P8.
The C/T column reports visible cores/threads.
}
\label{tab:benchmark1-platforms}
\footnotesize
\setlength{\tabcolsep}{2pt}
\renewcommand{\arraystretch}{1.06}
\begin{tabular}{>{\raggedright\arraybackslash}p{0.12\linewidth}
                >{\raggedright\arraybackslash}p{0.185\linewidth}
                >{\raggedright\arraybackslash}p{0.11\linewidth}
                >{\raggedright\arraybackslash}p{0.095\linewidth}
                >{\raggedleft\arraybackslash}p{0.033\linewidth}
                >{\raggedleft\arraybackslash}p{0.088\linewidth}
                >{\raggedright\arraybackslash}p{0.20\linewidth}
                >{\raggedright\arraybackslash}p{0.075\linewidth}}
\toprule
Host & Processor & Arch. & OS & C/T & Mem. & Cache summary & Comp.\\
\midrule
grice
  & Apple M3 Max & AArch64 & macOS 26.3 & 16 & 119 GiB
  & L1d 64 KiB; L2 4 MiB & g++ 14.3\\
cswaterloo
  & EPYC 9554, KVM/QEMU & x86-64 & Ubuntu 24.04 & 64v & 1024 GiB
  & L1d 4 MiB; L2 32 MiB; L3 1 GiB rep. & g++ 13.3\\
\bottomrule
\end{tabular}
\end{table}

%% file: tables/aarch64-cpp-ns-results.tex
\begin{table}[!t]
\centering
\caption{Times in nanoseconds per value on AArch64 (grice) for Benchmark~1.
\newline The values are means over five repetitions.
For the meanings of P$n$, see Table~\ref{tab:phases}.
}
\label{tab:aarch64-cpp-ns-results}
\small
\setlength{\tabcolsep}{3pt}
\begin{tabular}{lrrrrrrrr}
\toprule
$n$ & P1 & P2 & P3 & P4 & P5 & P6 & P7 & P8\\
\midrule
$10^6$ & 2.130 & 0.464 & 1.956 & 2.924
       & 0.864 & 0.598 & 1.332 & 1.334\\
$10^7$ & 5.709 & 0.465 & 2.219 & 4.857
       & 0.844 & 0.599 & 1.336 & 1.390\\
$10^8$ & 6.297 & 0.472 & 2.111 & 5.055
       & 0.844 & 0.594 & 1.297 & 1.339\\
$10^9$ & 6.767 & 0.466 & 2.115 & 5.322
       & 0.843 & 0.600 & 1.284 & 1.335\\
\bottomrule
\end{tabular}
\end{table}

%% file: tables/x86-cpp-ns-results.tex
\begin{table}[!t]
\centering
\caption{Times in nanoseconds per value on x86-64 (cswaterloo) for Benchmark~1.  The values are means over five repetitions.}
\label{tab:x86-cpp-ns-results}
\small
\setlength{\tabcolsep}{3pt}
\begin{tabular}{lrrrrrrrr}
\toprule
$n$ & P1 & P2 & P3 & P4 & P5 & P6 & P7 & P8\\
\midrule
$10^6$ & 12.258 & 0.826 & 4.788 & 7.534
       & 2.720 & 4.830 & 2.784 & 4.968\\
$10^7$ & 16.815 & 0.829 & 6.679 & 10.291
       & 2.704 & 4.752 & 2.769 & 4.984\\
$10^8$ & 19.731 & 0.827 & 7.835 & 11.714
       & 2.704 & 4.732 & 2.767 & 4.909\\
$10^9$ & 30.300 & 0.824 & 12.694 & 20.241
       & 2.705 & 4.729 & 2.770 & 4.860\\
\bottomrule
\end{tabular}
\end{table}

%% file: tables/cross-platform-ratios.tex
\begin{table}[!t]
\centering
\caption{Selected ratios at $10^9$ values for Benchmark 1.  
For the meanings of P$n$, see Table~\ref{tab:phases}.}
\label{tab:cross-platform-ratios}
\small
\setlength{\tabcolsep}{3pt}
\begin{tabular}{lrrrrrr}
\toprule
Host (architecture) & P1/P2 &P3/P5 & P4/P3 & P4/P5 & P4/P6 &  P8/P7\\
\midrule
grice (AArch64) & 14.51 & 2.51 & 2.52 & 6.32 & 8.87 & 1.04\\
cswaterloo (x86-64)  & 36.76 & 4.69 & 1.59 & 7.48 & 4.28 & 1.75\\
\bottomrule
\end{tabular}
\end{table}

%% file: tables/platform-rationale.tex
\begin{table}[!t]
\centering
\caption{Rationale for platform inclusion.  The table is intended to
show coverage of qualitatively different memory hierarchies and deployment
classes, not to rank processors.}
\label{tab:platform-rationale}
\footnotesize
\setlength{\tabcolsep}{3pt}
\renewcommand{\arraystretch}{1.08}
\begin{tabular}{>{\raggedright\arraybackslash}p{0.16\linewidth}
                >{\raggedright\arraybackslash}p{0.76\linewidth}}
\toprule
Host & Reason for inclusion\\
\midrule
\multicolumn{2}{l}{\textit{AArch64 hosts}}\\
rpiacn & Small AArch64 Linux board  with constrained memory; checks the embedded/single-board regime (Raspberry Pi).\\
seville & AArch64 Linux host with mixed Cortex-A55/A76-class cores and modest cache (Orange Pi).\\
grice & Current high-end Apple Silicon AArch64 laptop with large unified memory.\\
\midrule
\multicolumn{2}{l}{\textit{AMD x86-64 hosts}}\\
serwin & AMD Ryzen consumer x86-64 platform under Windows; complements the server-class x86 hosts.\\
cswaterloo & Large-memory AMD EPYC virtualized server with 57-bit virtual addresses reported.\\
\midrule
\multicolumn{2}{l}{\textit{Intel x86-64 hosts}}\\
giraffe & Older Intel Core desktop with small cache and memory; exposes memory-hierarchy sensitivity.\\
panamint & Intel Core i7-8086K Cygwin result; typical mid-age consumer platform.\\
triangle & Intel Xeon server-class machine; modern Intel server.\\
\bottomrule
\end{tabular}
\end{table}

%% file: tables/platforms.tex
\begin{table}[!t]
\centering
\caption{Benchmark~2 platforms.  These hosts were used for the broader
cross-host comparison of the generic-box benchmark.  The experiments are
single-threaded; the C/T column reports visible cores/threads where
available.  Cache entries are summaries of the reported data-cache
hierarchy. \newline$^*$ The panamint row is from one Cygwin run; the other rows summarize multi-run log bundles.}
\label{tab:platforms}
\footnotesize
\setlength{\tabcolsep}{3pt}
\renewcommand{\arraystretch}{1.08}
\begin{tabular}{>{\raggedright\arraybackslash}p{0.12\linewidth}
                >{\raggedright\arraybackslash}p{0.20\linewidth}
                >{\raggedright\arraybackslash}p{0.19\linewidth}
                >{\raggedright\arraybackslash}p{0.065\linewidth}
                >{\raggedright\arraybackslash}p{0.11\linewidth}
                >{\raggedright\arraybackslash}p{0.25\linewidth}}
\toprule
Host & Processor & OS/compiler & C/T & Mem. & Cache summary\\
\midrule
\multicolumn{5}{l}{\textit{AArch64 hosts}}\\
rpiacn
  & Raspberry~Pi~5,
  {\sc bcm}2712 Cortex-A76
  & Debian 13; \texttt{g++}~14.2
  & 4 
  & 7.9~GiB
  & L1d 256 KiB; L2 2 MiB; L3 2 MiB\\
seville
  & Orange~Pi~5, {\sc rk}3588{\sc s} Cortex-A55/A76
  & Ubuntu 22.04; \texttt{g++}~11.4
  & 8 
  &  15~GiB
  & L1d 384 KiB; L2 2.5 MiB; L3 3 MiB\\
grice
  & Apple M3 Max
  & macOS 26.3; \texttt{g++}~14.3
  & 16 
  &  119~GiB
  & L1d 64 KiB; L2 4 MiB\\
\midrule
\multicolumn{5}{l}{\textit{AMD x86-64 hosts}}\\
serwin
  & Ryzen 7 7735HS
  & Windows~11/ Cygwin; \texttt{g++}~13.4
  & 8/16 
  &  30~GiB
  & L1d 64 KiB/core; L2 512 KiB/core; L3 16 MiB\\
cswaterloo
  & EPYC 9554, KVM/QEMU
  & Ubuntu 24.04; \texttt{g++}~13.3
  & 64v 
  &  1024~GiB
  & L1d 4 MiB; L2 32 MiB; L3 1~GiB rep.\\
\midrule
\multicolumn{5}{l}{\textit{Intel x86-64 hosts}}\\
giraffe
  & Core i3-3220
  & Ubuntu 24.04.3; \texttt{g++}~13.3
  & 2/4 
  &  7.5~GiB
  & L1d 64 KiB; L2 512 KiB; L3 3 MiB\\
panamint
  & Core i7-8086K
  & Windows~11/ Cygwin; \texttt{g++}~16.0.1
  & 6/12 
  &  30~GiB
  & L3 12 MB\\
triangle
  & Xeon Silver 4314
  & Ubuntu 22.04.5; \texttt{g++}~11.4
  & 32/64 
  &  251~GiB
  & L1d 1.5 MiB; L2 40 MiB; L3 48 MiB\\
\bottomrule
\end{tabular}
\end{table}

%% file: tables/benchmark2-o3-crosshost.tex
\begin{table}[!t]
\centering
\caption{Generic-box benchmark at optimization \texttt{-O3} and $10^8$ values.
Each entry is a speedup ratio relative to the heap-header representation;
larger values mean that heap headers are slower.  H/Lo denotes
\textsc{HaBox}/\textsc{LoBox}, and H/Nan denotes
\textsc{HaBox}/\textsc{NanBox}.}
\label{tab:benchmark2-o3-crosshost}
\footnotesize
\setlength{\tabcolsep}{2.6pt}
\renewcommand{\arraystretch}{1.07}
\begin{tabular}{llrrrrrr}
\toprule
Host & Processor class
 & \multicolumn{2}{c}{tags}
 & \multicolumn{2}{c}{grouped \texttt+}
 & \multicolumn{2}{c}{boxed \texttt+}\\
\cmidrule(lr){3-4}\cmidrule(lr){5-6}\cmidrule(lr){7-8}
 & & H/Lo & H/Nan & H/Lo & H/Nan & H/Lo & H/Nan\\
\midrule

\multicolumn{8}{l}{\textit{AArch64 hosts}}\\
rpiacn     & ARM A76      & 20.37 & 11.83 & 2.17 & 1.94 & 3.98 & 3.46\\
seville    & ARM A55/A76   & 31.55 & 18.98 & 2.38 & 2.28 & 3.91 & 3.82\\
grice      & Apple M3 Max        & 10.10 &  9.27 & 1.89 & 1.68 & 4.86 & 3.61\\
\midrule
\multicolumn{8}{l}{\textit{AMD x86-64 hosts}}\\
serwin     & Ryzen 7             & 15.35 &  5.63 & 1.61 & 1.37 & 7.13 & 5.64\\
cswaterloo & EPYC                & 26.42 &  9.59 & 1.79 & 1.40 & 3.88 & 3.16\\
\midrule
\multicolumn{8}{l}{\textit{Intel x86-64 hosts}}\\
giraffe    & Core i3             &161.22 & 46.67 & 3.97 & 3.93 & 5.69 & 5.91\\
panamint & Core i7-8086K       &179.28 & 55.51 & 4.27 & 3.45 &12.81 &10.06\\
triangle   & Xeon Silver         & 20.52 &  6.34 & 3.20 & 2.66 & 5.04 & 4.47\\
\bottomrule
\end{tabular}
\end{table}

%% file: tables/benchmark2-summary-ranges.tex
\begin{table}[!t]
\centering
\caption{Summary of generic-box speedup ratios at optimization \texttt{-O3} and $10^8$
values.  H/Lo denotes \textsc{HaBox}/\textsc{LoBox}, and H/Nan denotes
\textsc{HaBox}/\textsc{NanBox}.  
``Main range'' excludes the old Core i3 and Core i7;
``Full range'' includes all platforms.
}
\label{tab:benchmark2-summary-ranges}
\footnotesize
\setlength{\tabcolsep}{3.5pt}
\renewcommand{\arraystretch}{1.08}
\begin{tabular}{lrrrr}
\toprule
Metric & Main range & Full range \\
\midrule
tags H/Lo      & 10.10--31.55 & 10.10--179.28\\
tags H/Nan     &  5.63--18.98 &  5.63--55.51\\
grouped\texttt{+} H/Lo &  1.61--3.20  &  1.61--4.27\\
grouped\texttt{+} H/Nan&  1.37--2.66  &  1.37--3.93\\
boxed\texttt{+} H/Lo   &  3.88--7.13  &  3.88--12.81\\
boxed\texttt{+} H/Nan  &  3.16--5.64  &  3.16--10.06\\
\bottomrule
\end{tabular}
\end{table}

%% file: tables/nan_lobox_ratios_O3_1e8.tex
\begin{table}[!t]
\centering
\caption{NanBox/LoBox timing ratios at optimization \texttt{-O3} and $10^8$ values.
Values greater than 1 mean that NaN-boxing is slower than low-bit tagging.}
\label{tab:nanbox-lobox-ratios-o3-1e8}
\footnotesize
\setlength{\tabcolsep}{4pt}
\begin{tabular}{lrrr}
\toprule
Host & tags & grouped \texttt{+} & boxed \texttt{+}\\
\midrule
\multicolumn{4}{l}{\textit{AArch64 hosts}}\\
rpiacn     & 1.72 & 1.12 & 1.15\\
seville    & 1.66 & 1.04 & 1.02\\
grice      & 1.09 & 1.12 & 1.35\\
\midrule
\multicolumn{4}{l}{\textit{AMD x86-64 hosts}}\\
serwin     & 2.73 & 1.18 & 1.26\\
cswaterloo & 2.75 & 1.28 & 1.23\\
\midrule
\multicolumn{4}{l}{\textit{Intel x86-64 hosts}}\\
giraffe    & 3.45 & 1.01 & 0.96\\
panamint   & 3.23 & 1.24 & 1.27\\
triangle   & 3.24 & 1.20 & 1.13\\
\bottomrule
\end{tabular}
\end{table}

%% file: tables/alt_form_instruction_counts_O3.tex
\begin{table}[!t]
\begin{center}
\footnotesize
\caption{Static fast-path instruction counts for forming boxed values
in the alternative-representation experiment at \texttt{-O3}.  Entries
of the form \texttt{new + n} exclude the allocator itself and count only
the visible caller-side instructions.}
\renewcommand{\tabcolsep}{5pt}
\begin{tabular}{lcc}
\toprule
Operation & x86-64 & AArch64 \\
\midrule
HaBox Nat/Int/Flo       & \texttt{new + 3} & \texttt{new + 2} \\
HaBox Cons              & \texttt{new + 4} & \texttt{new + 3} \\
LoBox Nat/Int           & 1 & 2 \\
LoBox Flo               & \texttt{new + 4} & \texttt{new + 3} \\
LoBox Cons              & \texttt{new + 5} & \texttt{new + 3} \\
NanBox Nat              & 1 & 1 \\
NanBox Int, checked     & 6 & 4 \\
NanBox Flo, non-qNaN    & 5 & 4 \\
NanBox Flo, imported qNaN & 5 & 4 \\
NanBox Cons             & \texttt{new + 7} & \texttt{new + 5} \\
\bottomrule
\end{tabular}
\label{tab:alt-form-instruction-counts-o3}
\end{center}
\end{table}

%% file: tables/alt_access_instruction_counts_O3.tex
\begin{table}[!t]
\begin{center}
\footnotesize
\caption{Static fast-path instruction counts for tag examination and
payload extraction in the alternative-representation experiment at
\texttt{-O3}.  Counts are for the visible inlined fast paths.}
\renewcommand{\tabcolsep}{5pt}
\begin{tabular}{lcc}
\toprule
Operation & x86-64 & AArch64 \\
\midrule
HaBox tag                         & 2 & 2 \\
LoBox tag                         & 2 & 2 \\
NanBox tag, boxed non-float path  & 9 & 6--7 \\[1ex]
HaBox Nat/Int/Flo payload         & 1 & 1 \\
LoBox Nat/Int payload             & 1 & 1 \\
LoBox Flo payload                 & 2 & 2 \\[1ex]
NanBox Nat payload                & 1 & 1 \\
NanBox Int payload                & 2 & 1 \\
NanBox Flo payload                & 1 & 1 \\[1ex]
HaBox car+cdr payloads            & 2 & 2 \\
LoBox car+cdr payloads            & 3 & 3 \\
NanBox car+cdr payloads           & 4 & 3 \\
\bottomrule
\end{tabular}
\label{tab:alt-access-instruction-counts-o3}
\end{center}
\end{table}